\documentclass[aps,pra,twocolumn,floatfix,superscriptaddress]{revtex4}

\usepackage{amssymb,amsmath,amstext}                
\usepackage{graphicx}                                               
\usepackage{epstopdf}                                               
\usepackage{color}                                                     
\usepackage{bm}                                                        
\usepackage{appendix}                                              
\usepackage[utf8]{inputenc}
\usepackage{bbold}
\usepackage{bbm}
\usepackage{ulem}
\usepackage{latexsym}
\usepackage[colorlinks=true,citecolor=blue,linkcolor=magenta]{hyperref}
\usepackage{mathtools}
\usepackage{mathtools}

\def\be{\begin{equation}}
\def\ee{\end{equation}}
\def\bea{\begin{eqnarray}}
\def\eea{\end{eqnarray}}

\def\bi{\begin{itemize}}
\def\ei{\end{itemize}}
\def\ben{\begin{enumerate}}
\def\een{\end{enumerate}}

\definecolor{dgreen} {RGB}{0,100,0}

\begin{document} 

\title{Lack of a genuine time crystal in a chiral soliton model}

\author{Andrzej Syrwid} 
\affiliation{
Instytut Fizyki Teoretycznej, 
Uniwersytet Jagiello\'nski, ulica Profesora Stanis\l{}awa \L{}ojasiewicza 11, PL-30-348 Krak\'ow, Poland}
\author{Arkadiusz Kosior} 
\affiliation{
Instytut Fizyki Teoretycznej, 
Uniwersytet Jagiello\'nski, ulica Profesora Stanis\l{}awa \L{}ojasiewicza 11, PL-30-348 Krak\'ow, Poland}
\affiliation{Max-Planck-Institut f\"ur Physik Komplexer Systeme,
N\"othnitzer Strasse 38, D-01187, Dresden, Germany}
\author{Krzysztof Sacha} 
\affiliation{
Instytut Fizyki Teoretycznej, 
Uniwersytet Jagiello\'nski, ulica Profesora Stanis\l{}awa \L{}ojasiewicza 11, PL-30-348 Krak\'ow, Poland}

\begin{abstract}
In a recent publication [Phys. Rev. Lett. {\bf 124}, 178902] \"Ohberg and Wright claim that in a chiral soliton model it is possible to realize a genuine time crystal which corresponds to a periodic evolution of an inhomogeneous probability density in the lowest energy state. We show that this result is incorrect and present a solution which possesses lower energy with the corresponding probability density that does not reveal any motion. It implies that the authors' conclusion that a genuine time crystal can exist in the system they consider is not true.
\end{abstract}

\date{\today}

\maketitle


\section{Introduction}
The idea of a quantum time crystal was proposed by Wilczek in 2012 \cite{Wilczek2012}. He considered attractively interacting bosons on a ring which formed a localized wave-packet (more precisely a bright soliton) and, in the presence of a magnetic-like flux, were supposed to move periodically along a ring even if the energy of the system was the lowest possible. 
The existence of such a genuine time crystal would involve spontaneous breaking of the continuous time translation symmetry into a discrete time translation symmetry in the system's ground state, in a full analogy to the spontaneous formation of ordinary space crystals \cite{Sacha2017rev}. It turned out that the system proposed by Wilczek was not a genuine time crystal, because in the limit of large number of bosons, the particle density corresponding to the ground state did not reveal any motion \cite{Bruno2013,Wilczek2013a,Syrwid2017}. It was also proven for a quite general class of systems with two-body interactions that a genuine time crystal cannot exist \cite{Bruno2013b,Watanabe2015,Watanabe2020}. Referring to systems with multi-particle interactions, Kozin and Kyriienko showed that a genuine time crystal could exist \cite{Kozin2019} but its experimental realization does not seem attainable \cite{khemani2020comment,KozinReply2020}. 
On the other hand,  another kind of time crystals has been demonstrated in the laboratory \cite{Zhang2017,Choi2017,Pal2018,Rovny2018,Smits2018}, i.e., the so-called discrete or Floquet time crystals where discrete time translation symmetry is spontaneously broken into another discrete time translation symmetry \cite{Sacha2015,Khemani16,ElseFTC}. Discrete time crystals and condensed matter physics in the time domain are becoming intensively developing research area \cite{Yao2017,Lazarides2017,Russomanno2017,
Zeng2017,Nakatsugawa2017,Ho2017,Huang2017,
Gong2017,Iemini2017,Wang2017,Flicker2018,Yu2018,
Tucker2018,Surace2018,Giergiel2018b,Lustig2018,
Giergiel2018c,Gambetta2018,Pizzi2019a,Pizzi2019,
Gambetta2019,Fan2019,Matus2019,Zhu2019,Buca2019a,
Lazarides2019,cai2019imaginary,
giergiel2020,kuros2020,kessler2020,
oberreiter2020,Russomanno2020,
Guo2013,Sacha15a,sacha16,Guo2016,
Guo2016a,Liang2017,Giergiel2017,
Mierzejewski2017,delande17,Kosior2018,
Kosior2018a,Giergiel2018,Giergiel2018a,
Mizuta2018,Bomantara2018}
(for comprehensive reviews see
\cite{Sacha2017rev,khemani2019brief,Guo2020})  but hunting for a genuine time crystal, which can be realized in the laboratory, is still continued.

Recently \"Ohberg and Wright have analyzed a mean-field description of a Bose system with a density-dependent gauge potential supporting chiral soliton solutions \cite{Ohberg2019}. They asserted that such a system could circumvent the no-go theorems \cite{Bruno2013b,Watanabe2015,Watanabe2020} and reveal a genuine time crystal behavior. In other words, the claim was that there existed a parameter regime where the lowest energy solution of the mean-field equation was a strongly localized soliton that evolved periodically along a ring~\cite{Ohberg2019}. The idea was very attractive because the considered system could be realized in ultra-cold atomic gases. However, it turned out that the energy of the system was not correctly calculated in Ref.~\cite{Ohberg2019} and the strongly localized soliton minimizes the energy when its velocity in the laboratory frame is zero \cite{SyrwidKosiorSacha2020}. 
In Ref.~\cite{ReplyOhbergWright2020} \"Ohberg and Wright admitted their error, but presented a new class of solutions by employing an ansatz that enforced the quantization of the soliton's velocity $u$ and argued that $u=0$ could be disallowed. Should the later be true, this would naturally allow for a genuine time crystal.

Here, we revisit the chiral soliton problem and show that the ansatz employed in Ref.~\cite{ReplyOhbergWright2020} restricts to a certain class of mean-field solutions only. 
There exist chiral soliton solutions that can move with any velocity
along a ring. Importantly,  a chiral soliton has the lowest energy when it does not move   and consequently no  genuine time crystal behavior is present in the system.

Before we switch to the chiral soliton problem analyzed in Refs.~\cite{Ohberg2019,ReplyOhbergWright2020} it is worth  presenting the basic arguments why a genuine time crystal cannot exist in a simpler solitonic problem, i.e., in the Wilczek model~\cite{Wilczek2012,Syrwid2017}.

\section{Wilczek model}
A single particle on a ring (whose position is denoted by an angle $\theta$) in the presence of a constant magnetic-like flux $\alpha$ is described by the Hamiltonian $H=(p-\alpha)^2/2$. The periodic boundary conditions on a ring, i.e. $\Psi(\theta+2\pi)=\Psi(\theta)$, imply the quantization of the particle momentum $p_n=n$ where $n$ is integer. If the flux $\alpha$ is not equal to an integer number, then in the ground state, $\Psi_{n}(\theta)=\mathrm{e}^{in\theta}/\sqrt{2\pi}$, the probability current is not zero,
\be
\frac{\partial H}{\partial p_n}=n-\alpha\ne 0,
\ee
where $n$ is the closest integer to $\alpha$. The corresponding probability density $|\Psi_n(\theta)|^2$ is spatially uniform and cannot be identified with a time crystal. Wilczek idea was to consider $N$ interacting  bosons on a ring in the presence of the constant magnetic-like flux $\alpha$ \cite{Wilczek2012}. If the interactions between particles are attractive and sufficiently strong it is known that the system forms a bright soliton in its lowest energy state. That is, in the solitonic regime, spontaneous breaking of the space translation symmetry occurs and the system's ground state collapses to a mean-field solution where all bosons occupy a bright soliton state \cite{Castin_LesHouches}. Wilczek hoped that in the presence of the flux $\alpha$, not only one could observe spontaneous breaking of the space translation symmetry, but also the soliton would move periodically on a ring. However, it does not happen and the easiest way to see it, is to analyze the center of mass of the system which is described by the Hamiltonian $H_{\rm CM}=(P-N\alpha)^2/(2N)$. The center of mass momentum is quantized, $P_n=n \in \mathbb{Z}$, but in the ground state, the probability current 
related to the center of mass motion vanishes in the $N\rightarrow\infty$ limit regardless of a choice of $\alpha$,
\be
\frac{\partial H_N}{\partial P_n}=\frac{n}{N}-\alpha\approx 0.
\label{CMmomentum}
\ee
Thus, if the $N$-particle system in the lowest energy state forms a bright soliton, then the soliton does not move when $N\rightarrow \infty$ \cite{Syrwid2017}. One might wonder whether the time crystal could be saved if we keep $N$ large but finite which, due to Eq.~(\ref{CMmomentum}), would correspond to a slowly moving ground state soliton solution.
 It turns out that we do need the infinite $N$ limit, because otherwise the center of mass position is subjected to quantum fluctuations and the mean-field bright soliton description breaks down. In the Wilczek's model, the quantum fluctuations of the center of mass position require infinite time to appear, only when $N\rightarrow\infty$ but $Ng=\rm constant$ (where $g$ is a contact interaction strength).
 Similarly, an ordinary space crystal is stable only in the thermodynamic limit ($N,V\rightarrow\infty$, $N/V=\rm constant$) where the energy difference between  symmetry broken states and the true ground state is infinitesimally small.  Otherwise, a space crystal would {\it melt} due to quantum fluctuations of the center of mass position \cite{Sacha2017rev}. 

It is worth analyzing the absence of a genuine time crystal also in the mean-field description. The mean-field approximation assumes that all $N$ bosons occupy the same single particle wave-function $\Psi(\theta,t)$ which fulfills the Gross-Pitaevskii equation (GPE) \cite{Pethick2002}. Assuming  dimensionless  variables as in Ref.~\cite{Ohberg2019}, the GPE reads
\be
i\partial_t\Psi=\left[ (-i\partial_\theta-\alpha)^2 +g|\Psi|^2\right]\Psi,
\label{GPE_Wilk}
\ee
with a contact interaction strength $g$,  a constant  $\alpha$ and $\langle\Psi|\Psi\rangle=1$. As the system is confined in a ring geometry  we assume that $\Psi$ fulfills periodic boundary conditions, $\Psi(\theta+2\pi ,t)=\Psi(\theta,t)$, and thus its phase can change only by $2\pi J$, where $J\in\mathbb{Z}$ is the phase winding number. The GPE, Eq.~(\ref{GPE_Wilk}), is generated by the action associated with the energy functional,
\be
E_{LAB}=\int\! d\theta\; \Psi^* \!\left[(-i\partial_\theta-\alpha)^2 +\frac{g}{2}|\Psi|^2\right]\!\Psi.
\label{elab_Wilk}
\ee
The energy $ E_{LAB}$ is the energy of the system in the laboratory frame which we want to minimize if we are looking for a genuine time crystal.
It turns out that for $g<-\pi$ and arbitrary $\alpha$, stable solitonic solutions of the GPE, Eq.~(\ref{GPE_Wilk}), exist and they can move with any velocity~$u$. These solutions are known analytically and can be expressed in terms of Jacobi elliptic functions and complete elliptic integrals~\cite{Carr2000,Kanamoto2003,Kanamoto2003B,Kanamoto2009}. 
Note that the fact that mean-field solitons on a ring can propagate with any velocity is consistent with the center of mass momentum quantization, i.e. in the limit $N\rightarrow \infty$ the momentum per particle $n/N$ becomes a continuous variable. Thus in contrast to the initial Wilczek's claim it does not matter if $\alpha$ is integer or not, the lowest energy state represented by a soliton solution reveals no periodic evolution.

\section{Chiral soliton model}
Let us consider the system of $N$ attractive bosons on a ring in the presence of  density-dependent gauge potential  \cite{Edmonds2013,Ohberg2019,ReplyOhbergWright2020}.  Within the mean-field description all bosons populate a Bose-Einstein condensate, where the condensate wave-function $\Psi(\theta,t)$ fulfills periodic boundary conditions, i.e., $\Psi(\theta+2\pi ,t)=\Psi(\theta,t)$.
In the dimensionless variables the laboratory frame energy per particle of the system reads 
\be
{\cal E}_{LAB}=\int\! d\theta\; \Psi^* \!\left[(-i\partial_\theta-A)^2+W+\frac{g}{2}|\Psi|^2\right]\!\Psi ,
\label{elab_CS}
\ee
where  $A=\frac{q}{2}+a|\Psi|^2$ is the density dependent vector potential, $W=\frac{q^2}{4}$ is a scalar potential, $q$  is an integer and $a$ determines the strength of the first-order density-dependent contribution to the vector potential. From now on we will refer to this model as a chiral soliton model. The chiral soliton model can be realized in ultracold atomic setups where the gauge fields $W$, $A$ arise as effective potentials due to light-matter interactions \cite{Edmonds2013}. In particular, $q$ is related to the gradient of the laser's phase and its quantization results from the winding number of the  Laguerre-Gaussian laser beam \cite{Ohberg2019}. The time evolution of the system is governed by the time-dependent GPE 
\be
i\partial_t\Psi=\left[(-i\partial_\theta-A)^2 -a j+ W+g|\Psi|^2\right]\Psi,
\label{GPE_W}
\ee
generated by the action associated with Eq.~(\ref{elab_CS}), where
\be
j=-i\Psi^*(\partial_\theta-iA)\Psi+c.c.
\label{probCur}
\ee
is the nonlinear current. In order to answer the question whether a genuine time crystal exists in this system we are going seek for the lowest energy solution in the frame moving with a velocity $u$. After that, we shall return to the laboratory frame and evaluate its energy. If the soliton has a  minimal energy for $u=0$, then no genuine time crystal exists. It is crucial to be in the soliton regime where formation of a localized wave-packet is energetically favorable because only non-homogeneous probability density that evolves periodically in time can represent a time crystal.

Switching to  the frame moving with a velocity $u$, $\Psi'(\theta,t)=\mathrm{e}^{ut \partial_\theta}\Psi(\theta,t)=\Psi(\theta+ut,t)$, the GPE  reads
\be
i\partial_t\Psi'=\left[ \left(-i\partial_\theta-A'\right)^2-a j'+(g-2au)|\Psi'|^2\right]\Psi',
\label{GPE}
\ee
where
\be
j'=-i\Psi'^*(\partial_\theta-iA')\Psi'+c.c.,
\ee
$A'=\frac{u}{2}+\frac{q}{2}+a|\Psi'|^2$ and constant contributions are accounted in the chemical potential. Note that in the moving frame the wavefunction $\Psi'$ also fulfills periodic boundary conditions, i.e., $\Psi'(\theta+2\pi,t)=\Psi'(\theta,t)$.

\begin{figure}
\includegraphics[width=1\columnwidth]{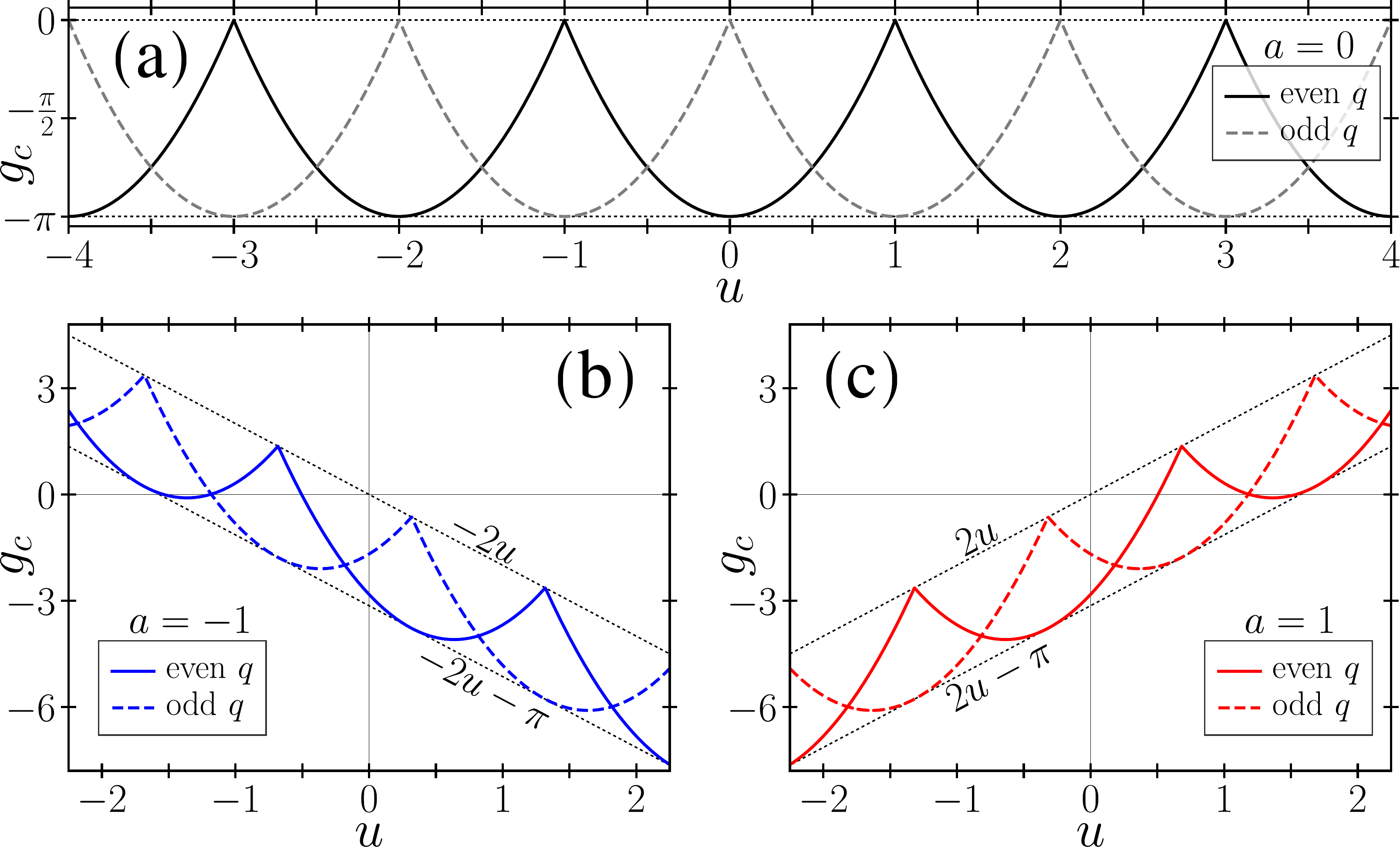} 
\caption{Critical interaction strength $g_c$, Eq.~(\ref{gc_val}), for the chiral soliton model. Panel (a)  illustrates the situation where no density-dependent gauge potential is present, i.e. when $a=0$, which corresponds to the Wilczek model with $\alpha=\frac{q}{2}$. An influence of a density-dependent gauge potential is shown in panels (b) and (c), where $a=-1$ and $a=+1$, respectively. Note that $-\pi+2au  \leq g_c \leq 2au$, which is indicated by dotted lines. In every panel the case of even (odd) $q$ is represented by solid (dashed) lines. 
  }
\label{fig_gc}   
\end{figure}

In order to find the lowest energy stationary solution $\Psi'(\theta)$ in the moving frame we evolve the GPE, Eq.~(\ref{GPE}), in the imaginary time \cite{Lehtovaara2007}. A uniform solution looses its stability for sufficiently strong interparticle attraction $g$, where formation of a localized lump -- a soliton -- becomes more energetically favourable.  To identify a parameter regime of a solitonic phase in the moving frame, we perform a Bogoliubov stability analysis of the uniform solution with the phase winding number $J\in\mathbb{Z}$, $\Psi'=\mathrm{e}^{i J \theta}/\sqrt{2\pi}$. 
That is, we study the linear stability of the stationary solution $\Psi'$ of the GPE in Eq.~(\ref{GPE}) under a small perturbation $\delta \Psi'$, i.e. $\Psi'\rightarrow \Psi_\delta'=\Psi' + \delta \Psi'$, where up to a trivial phase evolution
\be
\delta\Psi'(\theta,t)=
\displaystyle{\sum_{k\in \mathbb{Z}}\left(u_k\mathrm{e}^{ik\theta}\mathrm{e}^{-i\omega_k t}+v_k^*\mathrm{e}^{-ik\theta}\mathrm{e}^{i\omega_k^* t} \right)},
\ee
with $(u_k,v_k)$ and  $\omega_k$ being the eigenstates and eigenvalues of the Bogoliubov-de Gennes equations~\cite{Castin_LesHouches}  respectively. 
It is worth emphasizing that for each real eigenvalue $\omega_k$ corresponding to the eigenvector with a positive norm $\mathcal{N}_k=\big< u_k \big| u_k \big>-\big< v_k \big| v_k \big>=+1$ (the so-called {\it"+ family"} of the Bogoliubov modes~\cite{Castin_LesHouches}), there exists also an  eigenvalue $-\omega_k$ related to the eigenvector with a negative norm $\mathcal{N}_k=-1$ ({\it"$-$ family"}). By employing the Bogoliubov formalism one easily finds that for $g<g_d^{(J)}$, 
\be
g_d^{(J)}=-\pi+2a(2J-q)-3a^2/\pi,
\ee
the uniform solution $\Psi'=\mathrm{e}^{iJ\theta}/\sqrt{2\pi}$ is dynamically unstable. However, for $g_d^{(J)}<g<g_{c}^{(J)}$, where 
\be
g_c^{(J)} = -\pi+2au+4\pi\left(\Omega-J\right)^2,
\ee
with $\Omega=\frac{u}{2}+\frac{q}{2}+\frac{a}{2\pi}$, there is a negative eigenvalue of a {\it ``+ family''} Bogoliubov mode~\cite{Castin_LesHouches}, and consequently $\Psi'=\mathrm{e}^{iJ\theta}/\sqrt{2\pi}$ describes a dynamically stable excited state of the system.  In a result, for $g<g_c$,	
\be
g_c = \min_{J\in \mathbb{Z}} g_c^{(J)},
\label{gc_val}
\ee
there exists a stationary soliton solution which represents the lowest energy state of the system in the moving frame. 
 Note that the critical values of the interaction strength $g$ are different than reported in Refs.~\cite{Ohberg2019,ReplyOhbergWright2020}. In Fig.~\ref{fig_gc} we illustrate what is the influence of different values of the parameters $a$ and $q$ on  the critical interaction strength $g_c$, Eq.~(\ref{gc_val}), at which a chiral soliton moving with velocity $u$ appears \cite{q_footnote}.

\begin{figure} 	            
\includegraphics[width=1\columnwidth]{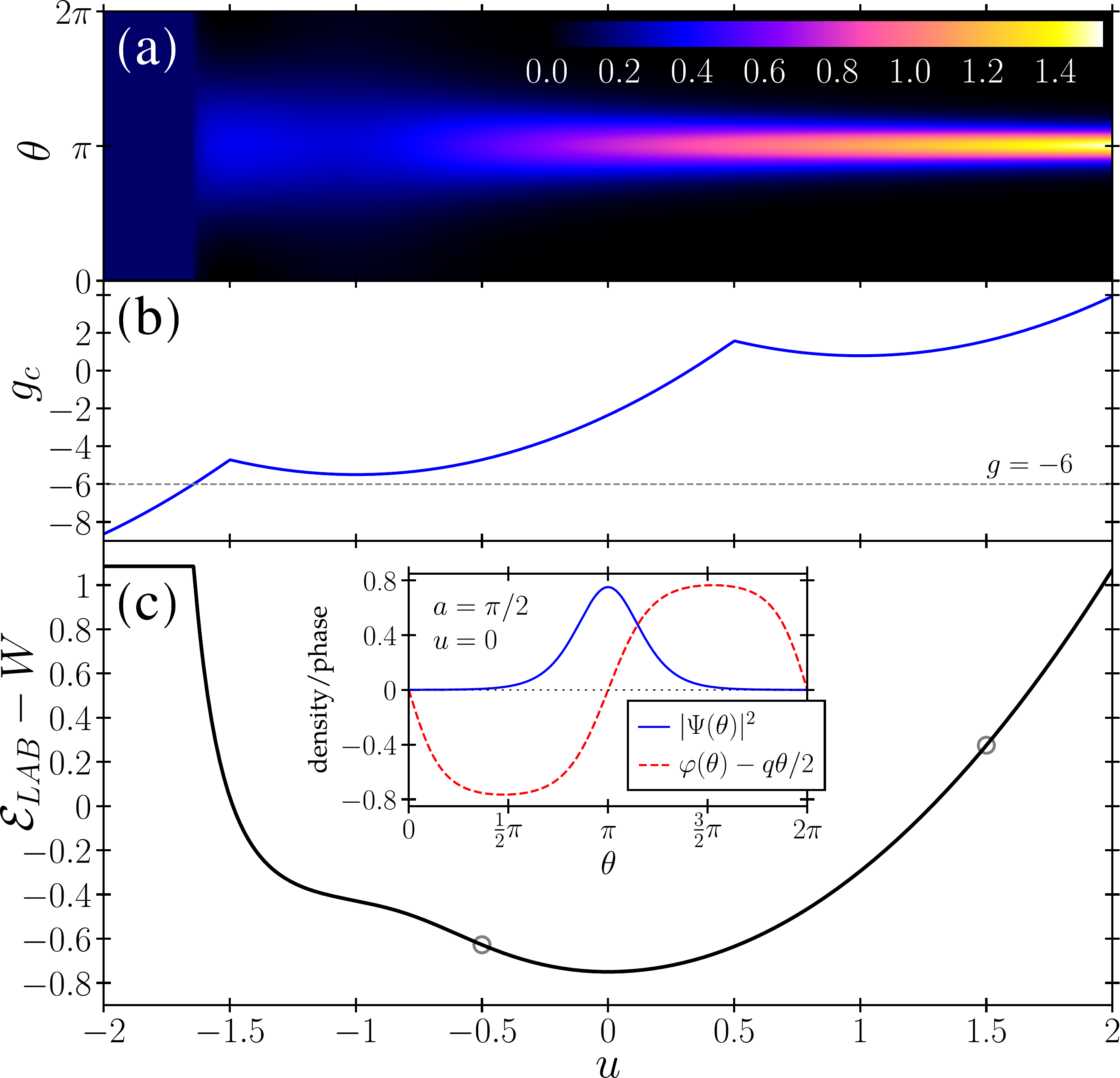}       
\caption{
Results of numerical simulations for exemplary parameters $a=\pi/2$, $g=-6$ and even $q$. (a) color coded plot of the chiral soliton density versus $u$. (b) critical value $g_c$ of interaction strength versus $u$ (dashed line shows $g=-6$). (c) the energy ${\cal E}_{LAB}$ (\ref{elab_CS}) of a chiral soliton moving with a velocity $u$. 
Circles indicate energies of solutions obtained within the ansatz used in Ref.~\cite{ReplyOhbergWright2020}, i.e. for $u=-0.5$ and $u=1.5$. Clearly, the ansatz solutions do not represent the system ground state.
Density $|\Psi(\theta)|^2$ and phase $\varphi(\theta)={\rm Arg}(\Psi)$ of the chiral soliton for  $u=0$ which minimizes $\mathcal{E}_{LAB}$ are depicted in the inset. 
  }
\label{fig}   
\end{figure}

Having the lowest energy soliton solution $\Psi'(\theta)$ in the frame moving with a velocity $u$, we return to the laboratory frame. This yields a solution moving periodically on a ring  $\Psi(\theta,t)=\mathrm{e}^{-ut \partial_\theta}\Psi'(\theta)=\Psi'(\theta-ut)$. 
In Fig.~\ref{fig} we present the results obtained for parameters for which \"Ohberg and Wright claim an existence of the time crystal, i.e. $a=\pi/2, \, g=-6$ and even $q$, but the final conclusion is the same for any choice of parameters $a,g$ and $q\in\mathbb{Z}$.
While the soliton solutions that fulfill  periodic boundary conditions exist for any $u\gtrsim-1.64$, the laboratory frame energy ${\cal E}_{LAB}$, Eq.~(\ref{elab_CS}), is minimal when $u=0$ and consequently in the lowest energy state no motion of the soliton is allowed. The corresponding density $|\Psi|^2$ and phase $\varphi={\rm Arg}(\Psi)$ of the lowest energy solution is depicted in the inset of Fig.~\ref{fig}~(c). We stress that a nontrivial phase $\varphi$ of the stationary solution visible in the inset is due to a nonzero $A$. Indeed, according to the continuity equation, the probability current $j=2|\Psi|^2(\partial_\theta \varphi -A)$ for stationary states can be nonzero but must be constant.

In Ref.~\cite{ReplyOhbergWright2020} where the existence of a genuine time crystal is claimed, the authors introduce the following ansatz,
\be
\Psi(\theta,t)=\mathrm{e}^{i\Theta(\theta,t)}\Phi(\theta-ut,t),
\label{AnsatzOW}
\ee
with
\be
\Theta(\theta,t)=\frac{q\theta}{2}+\frac{u\theta}{2}+a\int^\theta d\theta'|\Phi(\theta',t)|^2.
\ee
 Substitution of the ansatz to the GPE, Eq.~(\ref{GPE_W}), significantly simplifies the equation but in general implies twisted boundary conditions  $\Phi$
\be\label{twisted}
\Phi(\theta-ut+2\pi,t)= \mathrm{e}^{-i[\Theta(\theta+2\pi,t)-\Theta(\theta,t)]}\Phi(\theta-ut,t).
\ee
However, if one insists (as it is done in Ref.~\cite{ReplyOhbergWright2020}) that
\be
\Theta(\theta+2\pi,t)-\Theta(\theta,t)= 2\pi k, \quad k\in\mathbb{Z},
\ee
then $\Phi$ fulfills periodic boundary conditions (similarly as $\Psi$) which enforces that the velocity $u$ is allowed to take quantized values only --- e.g. for even $q$, the velocity $u=2k-a/\pi$. Then, the ansatz represents a certain class of solutions only. If it was a general solution, then velocity would be quantized even in the case of $a=0$ where analytical soliton solutions propagating with any $u$ are known~\cite{Carr2000,Kanamoto2003,Kanamoto2003B,Kanamoto2009}. Importantly, the ansatz does not describe the ground state of the system, see Fig.~\ref{fig}~(c). The latter corresponds to stationary probability density and consequently, the system does not represent a genuine time crystal.

\section{Discussion and conclusions}

 It turns out that it is not easy to realize a genuine time crystal. The initial proposition by Wilczek relied on attractively interacting bosons on a ring in the presence of a magnetic-like flux (the so-called Aharonov-Bohm ring)~\cite{Wilczek2012}. In the single particle case when the flux does not match the quantized values of the momentum of a particle on a ring, the probability current is non-zero even for the lowest energy state but the corresponding probability density is uniform and does not change over time. In a many-body case, the situation is quite opposite: although there exist spatially localized solutions which could travel non-dispersively, the ground state probability current is zero in the thermodynamic  limit \cite{Syrwid2017}.
 
\"Ohberg and Wright proposed an extension of the original Wilczek model, where they replaced a constant flux with a density-dependent gauge potential and claimed the existence of a time crystal behavior in the ground state of the system \cite{Ohberg2019}. The idea was very attractive because such a genuine time crystal could be realized in ultra-cold atoms laboratories. The publication triggered a debate in the literature whether the results are correct \cite{SyrwidKosiorSacha2020,ReplyOhbergWright2020}. 
In this work we are taking the final step of the discussion.

We have reexamined the chiral soliton model and showed that in Ref.~\cite{ReplyOhbergWright2020} a certain class of mean-field solutions is considered only and one can find other states which possess lower energy. It turns out that the ground state of the system is represented by a stationary probability density and consequently a genuine time crystal cannot be observed in the chiral soliton model.

\section*{Acknowledgements}
Support of the National Science Centre, Poland via Projects No.~2018/28/T/ST2/00372 (A.S.), No.~2016/21/B/ST2/01086 (A.K.) and No.~2018/31/B/ST2/00349
(K.S) is acknowledged. A.S. and A.K. acknowledge the support of the Foundation for Polish Science (FNP). 


\bibliography{ref_tc_book}

\begin{thebibliography}{80}
\expandafter\ifx\csname natexlab\endcsname\relax\def\natexlab#1{#1}\fi
\expandafter\ifx\csname bibnamefont\endcsname\relax
  \def\bibnamefont#1{#1}\fi
\expandafter\ifx\csname bibfnamefont\endcsname\relax
  \def\bibfnamefont#1{#1}\fi
\expandafter\ifx\csname citenamefont\endcsname\relax
  \def\citenamefont#1{#1}\fi
\expandafter\ifx\csname url\endcsname\relax
  \def\url#1{\texttt{#1}}\fi
\expandafter\ifx\csname urlprefix\endcsname\relax\def\urlprefix{URL }\fi
\providecommand{\bibinfo}[2]{#2}
\providecommand{\eprint}[2][]{\url{#2}}

\bibitem[{\citenamefont{Wilczek}(2012)}]{Wilczek2012}
\bibinfo{author}{\bibfnamefont{F.}~\bibnamefont{Wilczek}},
  \bibinfo{journal}{Phys. Rev. Lett.} \textbf{\bibinfo{volume}{109}},
  \bibinfo{pages}{160401} (\bibinfo{year}{2012}),
  \urlprefix\url{http://link.aps.org/doi/10.1103/PhysRevLett.109.160401}.

\bibitem[{\citenamefont{{Sacha} and {Zakrzewski}}(2018)}]{Sacha2017rev}
\bibinfo{author}{\bibfnamefont{K.}~\bibnamefont{{Sacha}}} \bibnamefont{and}
  \bibinfo{author}{\bibfnamefont{J.}~\bibnamefont{{Zakrzewski}}},
  \bibinfo{journal}{Rep. Prog. Phys.} \textbf{\bibinfo{volume}{81}},
  \bibinfo{pages}{016401} (\bibinfo{year}{2018}),
  \urlprefix\url{https://doi.org/10.1088/1361-6633/aa8b38}.

\bibitem[{\citenamefont{Bruno}(2013{\natexlab{a}})}]{Bruno2013}
\bibinfo{author}{\bibfnamefont{P.}~\bibnamefont{Bruno}},
  \bibinfo{journal}{Phys. Rev. Lett.} \textbf{\bibinfo{volume}{110}},
  \bibinfo{pages}{118901} (\bibinfo{year}{2013}{\natexlab{a}}),
  \urlprefix\url{http://link.aps.org/doi/10.1103/PhysRevLett.110.118901}.

\bibitem[{\citenamefont{Wilczek}(2013)}]{Wilczek2013a}
\bibinfo{author}{\bibfnamefont{F.}~\bibnamefont{Wilczek}},
  \bibinfo{journal}{Phys. Rev. Lett.} \textbf{\bibinfo{volume}{110}},
  \bibinfo{pages}{118902} (\bibinfo{year}{2013}),
  \urlprefix\url{http://link.aps.org/doi/10.1103/PhysRevLett.110.118902}.

\bibitem[{\citenamefont{Syrwid et~al.}(2017)\citenamefont{Syrwid, Zakrzewski,
  and Sacha}}]{Syrwid2017}
\bibinfo{author}{\bibfnamefont{A.}~\bibnamefont{Syrwid}},
  \bibinfo{author}{\bibfnamefont{J.}~\bibnamefont{Zakrzewski}},
  \bibnamefont{and} \bibinfo{author}{\bibfnamefont{K.}~\bibnamefont{Sacha}},
  \bibinfo{journal}{Phys. Rev. Lett.} \textbf{\bibinfo{volume}{119}},
  \bibinfo{pages}{250602} (\bibinfo{year}{2017}),
  \urlprefix\url{https://link.aps.org/doi/10.1103/PhysRevLett.119.250602}.

\bibitem[{\citenamefont{Bruno}(2013{\natexlab{b}})}]{Bruno2013b}
\bibinfo{author}{\bibfnamefont{P.}~\bibnamefont{Bruno}},
  \bibinfo{journal}{Phys. Rev. Lett.} \textbf{\bibinfo{volume}{111}},
  \bibinfo{pages}{070402} (\bibinfo{year}{2013}{\natexlab{b}}),
  \urlprefix\url{http://link.aps.org/doi/10.1103/PhysRevLett.111.070402}.

\bibitem[{\citenamefont{Watanabe and Oshikawa}(2015)}]{Watanabe2015}
\bibinfo{author}{\bibfnamefont{H.}~\bibnamefont{Watanabe}} \bibnamefont{and}
  \bibinfo{author}{\bibfnamefont{M.}~\bibnamefont{Oshikawa}},
  \bibinfo{journal}{Phys. Rev. Lett.} \textbf{\bibinfo{volume}{114}},
  \bibinfo{pages}{251603} (\bibinfo{year}{2015}),
  \urlprefix\url{http://link.aps.org/doi/10.1103/PhysRevLett.114.251603}.

\bibitem[{\citenamefont{Watanabe et~al.}(2020)\citenamefont{Watanabe, Oshikawa,
  and Koma}}]{Watanabe2020}
\bibinfo{author}{\bibfnamefont{H.}~\bibnamefont{Watanabe}},
  \bibinfo{author}{\bibfnamefont{M.}~\bibnamefont{Oshikawa}}, \bibnamefont{and}
  \bibinfo{author}{\bibfnamefont{T.}~\bibnamefont{Koma}},
  \bibinfo{journal}{Journal of Statistical Physics}
  \textbf{\bibinfo{volume}{178}}, \bibinfo{pages}{926} (\bibinfo{year}{2020}),
  ISSN \bibinfo{issn}{1572-9613},
  \urlprefix\url{https://doi.org/10.1007/s10955-019-02471-5}.

\bibitem[{\citenamefont{Kozin and Kyriienko}(2019)}]{Kozin2019}
\bibinfo{author}{\bibfnamefont{V.~K.} \bibnamefont{Kozin}} \bibnamefont{and}
  \bibinfo{author}{\bibfnamefont{O.}~\bibnamefont{Kyriienko}},
  \bibinfo{journal}{Phys. Rev. Lett.} \textbf{\bibinfo{volume}{123}},
  \bibinfo{pages}{210602} (\bibinfo{year}{2019}),
  \urlprefix\url{https://link.aps.org/doi/10.1103/PhysRevLett.123.210602}.

\bibitem[{\citenamefont{Khemani et~al.}(2020)\citenamefont{Khemani, Moessner,
  and Sondhi}}]{khemani2020comment}
\bibinfo{author}{\bibfnamefont{V.}~\bibnamefont{Khemani}},
  \bibinfo{author}{\bibfnamefont{R.}~\bibnamefont{Moessner}}, \bibnamefont{and}
  \bibinfo{author}{\bibfnamefont{S.~L.} \bibnamefont{Sondhi}},
  \bibinfo{eid}{arXiv:2001.11037} (\bibinfo{year}{2020}).

\bibitem[{\citenamefont{{Kozin} and {Kyriienko}}(2020)}]{KozinReply2020}
\bibinfo{author}{\bibfnamefont{V.~K.} \bibnamefont{{Kozin}}} \bibnamefont{and}
  \bibinfo{author}{\bibfnamefont{O.}~\bibnamefont{{Kyriienko}}},
  \bibinfo{journal}{arXiv e-prints} \bibinfo{eid}{arXiv:2005.06321}
  (\bibinfo{year}{2020}).

\bibitem[{\citenamefont{Zhang et~al.}(2017)\citenamefont{Zhang, Hess,
  Kyprianidis, Becker, Lee, Smith, Pagano, Potirniche, Potter, Vishwanath
  et~al.}}]{Zhang2017}
\bibinfo{author}{\bibfnamefont{J.}~\bibnamefont{Zhang}},
  \bibinfo{author}{\bibfnamefont{P.~W.} \bibnamefont{Hess}},
  \bibinfo{author}{\bibfnamefont{A.}~\bibnamefont{Kyprianidis}},
  \bibinfo{author}{\bibfnamefont{P.}~\bibnamefont{Becker}},
  \bibinfo{author}{\bibfnamefont{A.}~\bibnamefont{Lee}},
  \bibinfo{author}{\bibfnamefont{J.}~\bibnamefont{Smith}},
  \bibinfo{author}{\bibfnamefont{G.}~\bibnamefont{Pagano}},
  \bibinfo{author}{\bibfnamefont{I.-D.} \bibnamefont{Potirniche}},
  \bibinfo{author}{\bibfnamefont{A.~C.} \bibnamefont{Potter}},
  \bibinfo{author}{\bibfnamefont{A.}~\bibnamefont{Vishwanath}},
  \bibnamefont{et~al.}, \bibinfo{journal}{Nature}
  \textbf{\bibinfo{volume}{543}}, \bibinfo{pages}{217} (\bibinfo{year}{2017}),
  ISSN \bibinfo{issn}{0028-0836}, \bibinfo{note}{letter},
  \urlprefix\url{http://dx.doi.org/10.1038/nature21413}.

\bibitem[{\citenamefont{Choi et~al.}(2017)\citenamefont{Choi, Choi, Landig,
  Kucsko, Zhou, Isoya, Jelezko, Onoda, Sumiya, Khemani et~al.}}]{Choi2017}
\bibinfo{author}{\bibfnamefont{S.}~\bibnamefont{Choi}},
  \bibinfo{author}{\bibfnamefont{J.}~\bibnamefont{Choi}},
  \bibinfo{author}{\bibfnamefont{R.}~\bibnamefont{Landig}},
  \bibinfo{author}{\bibfnamefont{G.}~\bibnamefont{Kucsko}},
  \bibinfo{author}{\bibfnamefont{H.}~\bibnamefont{Zhou}},
  \bibinfo{author}{\bibfnamefont{J.}~\bibnamefont{Isoya}},
  \bibinfo{author}{\bibfnamefont{F.}~\bibnamefont{Jelezko}},
  \bibinfo{author}{\bibfnamefont{S.}~\bibnamefont{Onoda}},
  \bibinfo{author}{\bibfnamefont{H.}~\bibnamefont{Sumiya}},
  \bibinfo{author}{\bibfnamefont{V.}~\bibnamefont{Khemani}},
  \bibnamefont{et~al.}, \bibinfo{journal}{Nature}
  \textbf{\bibinfo{volume}{543}}, \bibinfo{pages}{221} (\bibinfo{year}{2017}),
  ISSN \bibinfo{issn}{0028-0836}, \bibinfo{note}{letter},
  \urlprefix\url{http://dx.doi.org/10.1038/nature21426}.

\bibitem[{\citenamefont{Pal et~al.}(2018)\citenamefont{Pal, Nishad, Mahesh, and
  Sreejith}}]{Pal2018}
\bibinfo{author}{\bibfnamefont{S.}~\bibnamefont{Pal}},
  \bibinfo{author}{\bibfnamefont{N.}~\bibnamefont{Nishad}},
  \bibinfo{author}{\bibfnamefont{T.~S.} \bibnamefont{Mahesh}},
  \bibnamefont{and} \bibinfo{author}{\bibfnamefont{G.~J.}
  \bibnamefont{Sreejith}}, \bibinfo{journal}{Phys. Rev. Lett.}
  \textbf{\bibinfo{volume}{120}}, \bibinfo{pages}{180602}
  (\bibinfo{year}{2018}),
  \urlprefix\url{https://link.aps.org/doi/10.1103/PhysRevLett.120.180602}.

\bibitem[{\citenamefont{Rovny et~al.}(2018)\citenamefont{Rovny, Blum, and
  Barrett}}]{Rovny2018}
\bibinfo{author}{\bibfnamefont{J.}~\bibnamefont{Rovny}},
  \bibinfo{author}{\bibfnamefont{R.~L.} \bibnamefont{Blum}}, \bibnamefont{and}
  \bibinfo{author}{\bibfnamefont{S.~E.} \bibnamefont{Barrett}},
  \bibinfo{journal}{Phys. Rev. Lett.} \textbf{\bibinfo{volume}{120}},
  \bibinfo{pages}{180603} (\bibinfo{year}{2018}),
  \urlprefix\url{https://link.aps.org/doi/10.1103/PhysRevLett.120.180603}.

\bibitem[{\citenamefont{Smits et~al.}(2018)\citenamefont{Smits, Liao, Stoof,
  and van~der Straten}}]{Smits2018}
\bibinfo{author}{\bibfnamefont{J.}~\bibnamefont{Smits}},
  \bibinfo{author}{\bibfnamefont{L.}~\bibnamefont{Liao}},
  \bibinfo{author}{\bibfnamefont{H.~T.~C.} \bibnamefont{Stoof}},
  \bibnamefont{and} \bibinfo{author}{\bibfnamefont{P.}~\bibnamefont{van~der
  Straten}}, \bibinfo{journal}{Phys. Rev. Lett.}
  \textbf{\bibinfo{volume}{121}}, \bibinfo{pages}{185301}
  (\bibinfo{year}{2018}),
  \urlprefix\url{https://link.aps.org/doi/10.1103/PhysRevLett.121.185301}.

\bibitem[{\citenamefont{Sacha}(2015{\natexlab{a}})}]{Sacha2015}
\bibinfo{author}{\bibfnamefont{K.}~\bibnamefont{Sacha}},
  \bibinfo{journal}{Phys. Rev. A} \textbf{\bibinfo{volume}{91}},
  \bibinfo{pages}{033617} (\bibinfo{year}{2015}{\natexlab{a}}),
  \urlprefix\url{http://link.aps.org/doi/10.1103/PhysRevA.91.033617}.

\bibitem[{\citenamefont{Khemani et~al.}(2016)\citenamefont{Khemani, Lazarides,
  Moessner, and Sondhi}}]{Khemani16}
\bibinfo{author}{\bibfnamefont{V.}~\bibnamefont{Khemani}},
  \bibinfo{author}{\bibfnamefont{A.}~\bibnamefont{Lazarides}},
  \bibinfo{author}{\bibfnamefont{R.}~\bibnamefont{Moessner}}, \bibnamefont{and}
  \bibinfo{author}{\bibfnamefont{S.~L.} \bibnamefont{Sondhi}},
  \bibinfo{journal}{Phys. Rev. Lett.} \textbf{\bibinfo{volume}{116}},
  \bibinfo{pages}{250401} (\bibinfo{year}{2016}),
  \urlprefix\url{http://link.aps.org/doi/10.1103/PhysRevLett.116.250401}.

\bibitem[{\citenamefont{Else et~al.}(2016)\citenamefont{Else, Bauer, and
  Nayak}}]{ElseFTC}
\bibinfo{author}{\bibfnamefont{D.~V.} \bibnamefont{Else}},
  \bibinfo{author}{\bibfnamefont{B.}~\bibnamefont{Bauer}}, \bibnamefont{and}
  \bibinfo{author}{\bibfnamefont{C.}~\bibnamefont{Nayak}},
  \bibinfo{journal}{Phys. Rev. Lett.} \textbf{\bibinfo{volume}{117}},
  \bibinfo{pages}{090402} (\bibinfo{year}{2016}),
  \urlprefix\url{http://link.aps.org/doi/10.1103/PhysRevLett.117.090402}.

\bibitem[{\citenamefont{Yao et~al.}(2017)\citenamefont{Yao, Potter, Potirniche,
  and Vishwanath}}]{Yao2017}
\bibinfo{author}{\bibfnamefont{N.~Y.} \bibnamefont{Yao}},
  \bibinfo{author}{\bibfnamefont{A.~C.} \bibnamefont{Potter}},
  \bibinfo{author}{\bibfnamefont{I.-D.} \bibnamefont{Potirniche}},
  \bibnamefont{and}
  \bibinfo{author}{\bibfnamefont{A.}~\bibnamefont{Vishwanath}},
  \bibinfo{journal}{Phys. Rev. Lett.} \textbf{\bibinfo{volume}{118}},
  \bibinfo{pages}{030401} (\bibinfo{year}{2017}),
  \urlprefix\url{http://link.aps.org/doi/10.1103/PhysRevLett.118.030401}.

\bibitem[{\citenamefont{Lazarides and Moessner}(2017)}]{Lazarides2017}
\bibinfo{author}{\bibfnamefont{A.}~\bibnamefont{Lazarides}} \bibnamefont{and}
  \bibinfo{author}{\bibfnamefont{R.}~\bibnamefont{Moessner}},
  \bibinfo{journal}{Phys. Rev. B} \textbf{\bibinfo{volume}{95}},
  \bibinfo{pages}{195135} (\bibinfo{year}{2017}),
  \urlprefix\url{https://link.aps.org/doi/10.1103/PhysRevB.95.195135}.

\bibitem[{\citenamefont{Russomanno et~al.}(2017)\citenamefont{Russomanno,
  Iemini, Dalmonte, and Fazio}}]{Russomanno2017}
\bibinfo{author}{\bibfnamefont{A.}~\bibnamefont{Russomanno}},
  \bibinfo{author}{\bibfnamefont{F.}~\bibnamefont{Iemini}},
  \bibinfo{author}{\bibfnamefont{M.}~\bibnamefont{Dalmonte}}, \bibnamefont{and}
  \bibinfo{author}{\bibfnamefont{R.}~\bibnamefont{Fazio}},
  \bibinfo{journal}{Phys. Rev. B} \textbf{\bibinfo{volume}{95}},
  \bibinfo{pages}{214307} (\bibinfo{year}{2017}),
  \urlprefix\url{https://link.aps.org/doi/10.1103/PhysRevB.95.214307}.

\bibitem[{\citenamefont{Zeng and Sheng}(2017)}]{Zeng2017}
\bibinfo{author}{\bibfnamefont{T.-S.} \bibnamefont{Zeng}} \bibnamefont{and}
  \bibinfo{author}{\bibfnamefont{D.~N.} \bibnamefont{Sheng}},
  \bibinfo{journal}{Phys. Rev. B} \textbf{\bibinfo{volume}{96}},
  \bibinfo{pages}{094202} (\bibinfo{year}{2017}),
  \urlprefix\url{https://link.aps.org/doi/10.1103/PhysRevB.96.094202}.

\bibitem[{\citenamefont{Nakatsugawa et~al.}(2017)\citenamefont{Nakatsugawa,
  Fujii, and Tanda}}]{Nakatsugawa2017}
\bibinfo{author}{\bibfnamefont{K.}~\bibnamefont{Nakatsugawa}},
  \bibinfo{author}{\bibfnamefont{T.}~\bibnamefont{Fujii}}, \bibnamefont{and}
  \bibinfo{author}{\bibfnamefont{S.}~\bibnamefont{Tanda}},
  \bibinfo{journal}{Phys. Rev. B} \textbf{\bibinfo{volume}{96}},
  \bibinfo{pages}{094308} (\bibinfo{year}{2017}),
  \urlprefix\url{https://link.aps.org/doi/10.1103/PhysRevB.96.094308}.

\bibitem[{\citenamefont{Ho et~al.}(2017)\citenamefont{Ho, Choi, Lukin, and
  Abanin}}]{Ho2017}
\bibinfo{author}{\bibfnamefont{W.~W.} \bibnamefont{Ho}},
  \bibinfo{author}{\bibfnamefont{S.}~\bibnamefont{Choi}},
  \bibinfo{author}{\bibfnamefont{M.~D.} \bibnamefont{Lukin}}, \bibnamefont{and}
  \bibinfo{author}{\bibfnamefont{D.~A.} \bibnamefont{Abanin}},
  \bibinfo{journal}{Phys. Rev. Lett.} \textbf{\bibinfo{volume}{119}},
  \bibinfo{pages}{010602} (\bibinfo{year}{2017}),
  \urlprefix\url{https://link.aps.org/doi/10.1103/PhysRevLett.119.010602}.

\bibitem[{\citenamefont{Huang et~al.}(2018)\citenamefont{Huang, Wu, and
  Liu}}]{Huang2017}
\bibinfo{author}{\bibfnamefont{B.}~\bibnamefont{Huang}},
  \bibinfo{author}{\bibfnamefont{Y.-H.} \bibnamefont{Wu}}, \bibnamefont{and}
  \bibinfo{author}{\bibfnamefont{W.~V.} \bibnamefont{Liu}},
  \bibinfo{journal}{Phys. Rev. Lett.} \textbf{\bibinfo{volume}{120}},
  \bibinfo{pages}{110603} (\bibinfo{year}{2018}),
  \urlprefix\url{https://link.aps.org/doi/10.1103/PhysRevLett.120.110603}.

\bibitem[{\citenamefont{Gong et~al.}(2018)\citenamefont{Gong, Hamazaki, and
  Ueda}}]{Gong2017}
\bibinfo{author}{\bibfnamefont{Z.}~\bibnamefont{Gong}},
  \bibinfo{author}{\bibfnamefont{R.}~\bibnamefont{Hamazaki}}, \bibnamefont{and}
  \bibinfo{author}{\bibfnamefont{M.}~\bibnamefont{Ueda}},
  \bibinfo{journal}{Phys. Rev. Lett.} \textbf{\bibinfo{volume}{120}},
  \bibinfo{pages}{040404} (\bibinfo{year}{2018}),
  \urlprefix\url{https://link.aps.org/doi/10.1103/PhysRevLett.120.040404}.

\bibitem[{\citenamefont{Iemini et~al.}(2018)\citenamefont{Iemini, Russomanno,
  Keeling, Schir\`o, Dalmonte, and Fazio}}]{Iemini2017}
\bibinfo{author}{\bibfnamefont{F.}~\bibnamefont{Iemini}},
  \bibinfo{author}{\bibfnamefont{A.}~\bibnamefont{Russomanno}},
  \bibinfo{author}{\bibfnamefont{J.}~\bibnamefont{Keeling}},
  \bibinfo{author}{\bibfnamefont{M.}~\bibnamefont{Schir\`o}},
  \bibinfo{author}{\bibfnamefont{M.}~\bibnamefont{Dalmonte}}, \bibnamefont{and}
  \bibinfo{author}{\bibfnamefont{R.}~\bibnamefont{Fazio}},
  \bibinfo{journal}{Phys. Rev. Lett.} \textbf{\bibinfo{volume}{121}},
  \bibinfo{pages}{035301} (\bibinfo{year}{2018}),
  \urlprefix\url{https://link.aps.org/doi/10.1103/PhysRevLett.121.035301}.

\bibitem[{\citenamefont{Wang et~al.}(2018)\citenamefont{Wang, Xing, Carlo, and
  Poletti}}]{Wang2017}
\bibinfo{author}{\bibfnamefont{R.~R.~W.} \bibnamefont{Wang}},
  \bibinfo{author}{\bibfnamefont{B.}~\bibnamefont{Xing}},
  \bibinfo{author}{\bibfnamefont{G.~G.} \bibnamefont{Carlo}}, \bibnamefont{and}
  \bibinfo{author}{\bibfnamefont{D.}~\bibnamefont{Poletti}},
  \bibinfo{journal}{Phys. Rev. E} \textbf{\bibinfo{volume}{97}},
  \bibinfo{pages}{020202(R)} (\bibinfo{year}{2018}),
  \urlprefix\url{https://link.aps.org/doi/10.1103/PhysRevE.97.020202}.

\bibitem[{\citenamefont{Flicker}(2018)}]{Flicker2018}
\bibinfo{author}{\bibfnamefont{F.}~\bibnamefont{Flicker}},
  \bibinfo{journal}{SciPost Phys.} \textbf{\bibinfo{volume}{5}},
  \bibinfo{pages}{1} (\bibinfo{year}{2018}),
  \urlprefix\url{https://scipost.org/10.21468/SciPostPhys.5.1.001}.

\bibitem[{\citenamefont{Yu et~al.}(2019)\citenamefont{Yu, Tangpanitanon,
  Glaetzle, Jaksch, and Angelakis}}]{Yu2018}
\bibinfo{author}{\bibfnamefont{W.~C.} \bibnamefont{Yu}},
  \bibinfo{author}{\bibfnamefont{J.}~\bibnamefont{Tangpanitanon}},
  \bibinfo{author}{\bibfnamefont{A.~W.} \bibnamefont{Glaetzle}},
  \bibinfo{author}{\bibfnamefont{D.}~\bibnamefont{Jaksch}}, \bibnamefont{and}
  \bibinfo{author}{\bibfnamefont{D.~G.} \bibnamefont{Angelakis}},
  \bibinfo{journal}{Phys. Rev. A} \textbf{\bibinfo{volume}{99}},
  \bibinfo{pages}{033618} (\bibinfo{year}{2019}),
  \urlprefix\url{https://link.aps.org/doi/10.1103/PhysRevA.99.033618}.

\bibitem[{\citenamefont{{Tucker} et~al.}(2018)\citenamefont{{Tucker}, {Zhu},
  {Lewis-Swan}, {Marino}, {Jimenez}, {Restrepo}, and {Rey}}}]{Tucker2018}
\bibinfo{author}{\bibfnamefont{K.}~\bibnamefont{{Tucker}}},
  \bibinfo{author}{\bibfnamefont{B.}~\bibnamefont{{Zhu}}},
  \bibinfo{author}{\bibfnamefont{R.~J.} \bibnamefont{{Lewis-Swan}}},
  \bibinfo{author}{\bibfnamefont{J.}~\bibnamefont{{Marino}}},
  \bibinfo{author}{\bibfnamefont{F.}~\bibnamefont{{Jimenez}}},
  \bibinfo{author}{\bibfnamefont{J.~G.} \bibnamefont{{Restrepo}}},
  \bibnamefont{and} \bibinfo{author}{\bibfnamefont{A.~M.} \bibnamefont{{Rey}}},
  \bibinfo{journal}{ArXiv e-prints} \bibinfo{eid}{arXiv:1805.03343}
  (\bibinfo{year}{2018}).

\bibitem[{\citenamefont{Surace et~al.}(2019)\citenamefont{Surace, Russomanno,
  Dalmonte, Silva, Fazio, and Iemini}}]{Surace2018}
\bibinfo{author}{\bibfnamefont{F.~M.} \bibnamefont{Surace}},
  \bibinfo{author}{\bibfnamefont{A.}~\bibnamefont{Russomanno}},
  \bibinfo{author}{\bibfnamefont{M.}~\bibnamefont{Dalmonte}},
  \bibinfo{author}{\bibfnamefont{A.}~\bibnamefont{Silva}},
  \bibinfo{author}{\bibfnamefont{R.}~\bibnamefont{Fazio}}, \bibnamefont{and}
  \bibinfo{author}{\bibfnamefont{F.}~\bibnamefont{Iemini}},
  \bibinfo{journal}{Phys. Rev. B} \textbf{\bibinfo{volume}{99}},
  \bibinfo{pages}{104303} (\bibinfo{year}{2019}),
  \urlprefix\url{https://link.aps.org/doi/10.1103/PhysRevB.99.104303}.

\bibitem[{\citenamefont{Giergiel
  et~al.}(2019{\natexlab{a}})\citenamefont{Giergiel, Dauphin, Lewenstein,
  Zakrzewski, and Sacha}}]{Giergiel2018b}
\bibinfo{author}{\bibfnamefont{K.}~\bibnamefont{Giergiel}},
  \bibinfo{author}{\bibfnamefont{A.}~\bibnamefont{Dauphin}},
  \bibinfo{author}{\bibfnamefont{M.}~\bibnamefont{Lewenstein}},
  \bibinfo{author}{\bibfnamefont{J.}~\bibnamefont{Zakrzewski}},
  \bibnamefont{and} \bibinfo{author}{\bibfnamefont{K.}~\bibnamefont{Sacha}},
  \bibinfo{journal}{New Journal of Physics} \textbf{\bibinfo{volume}{21}},
  \bibinfo{pages}{052003} (\bibinfo{year}{2019}{\natexlab{a}}),
  \urlprefix\url{https://doi.org/10.1088/1367-2630/ab1e5f}.

\bibitem[{\citenamefont{Lustig et~al.}(2018)\citenamefont{Lustig, Sharabi, and
  Segev}}]{Lustig2018}
\bibinfo{author}{\bibfnamefont{E.}~\bibnamefont{Lustig}},
  \bibinfo{author}{\bibfnamefont{Y.}~\bibnamefont{Sharabi}}, \bibnamefont{and}
  \bibinfo{author}{\bibfnamefont{M.}~\bibnamefont{Segev}},
  \bibinfo{journal}{Optica} \textbf{\bibinfo{volume}{5}}, \bibinfo{pages}{1390}
  (\bibinfo{year}{2018}),
  \urlprefix\url{http://www.osapublishing.org/optica/abstract.cfm?URI=optica-5-11-1390}.

\bibitem[{\citenamefont{Giergiel
  et~al.}(2019{\natexlab{b}})\citenamefont{Giergiel,
  Kuro\ifmmode~\acute{s}\else \'{s}\fi{}, and Sacha}}]{Giergiel2018c}
\bibinfo{author}{\bibfnamefont{K.}~\bibnamefont{Giergiel}},
  \bibinfo{author}{\bibfnamefont{A.}~\bibnamefont{Kuro\ifmmode~\acute{s}\else
  \'{s}\fi{}}}, \bibnamefont{and}
  \bibinfo{author}{\bibfnamefont{K.}~\bibnamefont{Sacha}},
  \bibinfo{journal}{Phys. Rev. B} \textbf{\bibinfo{volume}{99}},
  \bibinfo{pages}{220303(R)} (\bibinfo{year}{2019}{\natexlab{b}}),
  \urlprefix\url{https://link.aps.org/doi/10.1103/PhysRevB.99.220303}.

\bibitem[{\citenamefont{Gambetta
  et~al.}(2019{\natexlab{a}})\citenamefont{Gambetta, Carollo, Marcuzzi,
  Garrahan, and Lesanovsky}}]{Gambetta2018}
\bibinfo{author}{\bibfnamefont{F.~M.} \bibnamefont{Gambetta}},
  \bibinfo{author}{\bibfnamefont{F.}~\bibnamefont{Carollo}},
  \bibinfo{author}{\bibfnamefont{M.}~\bibnamefont{Marcuzzi}},
  \bibinfo{author}{\bibfnamefont{J.~P.} \bibnamefont{Garrahan}},
  \bibnamefont{and}
  \bibinfo{author}{\bibfnamefont{I.}~\bibnamefont{Lesanovsky}},
  \bibinfo{journal}{Phys. Rev. Lett.} \textbf{\bibinfo{volume}{122}},
  \bibinfo{pages}{015701} (\bibinfo{year}{2019}{\natexlab{a}}),
  \urlprefix\url{https://link.aps.org/doi/10.1103/PhysRevLett.122.015701}.

\bibitem[{\citenamefont{Pizzi et~al.}(2019)\citenamefont{Pizzi, Knolle, and
  Nunnenkamp}}]{Pizzi2019a}
\bibinfo{author}{\bibfnamefont{A.}~\bibnamefont{Pizzi}},
  \bibinfo{author}{\bibfnamefont{J.}~\bibnamefont{Knolle}}, \bibnamefont{and}
  \bibinfo{author}{\bibfnamefont{A.}~\bibnamefont{Nunnenkamp}},
  \bibinfo{journal}{Phys. Rev. Lett.} \textbf{\bibinfo{volume}{123}},
  \bibinfo{pages}{150601} (\bibinfo{year}{2019}),
  \urlprefix\url{https://link.aps.org/doi/10.1103/PhysRevLett.123.150601}.

\bibitem[{\citenamefont{{Pizzi} et~al.}(2019)\citenamefont{{Pizzi}, {Knolle},
  and {Nunnenkamp}}}]{Pizzi2019}
\bibinfo{author}{\bibfnamefont{A.}~\bibnamefont{{Pizzi}}},
  \bibinfo{author}{\bibfnamefont{J.}~\bibnamefont{{Knolle}}}, \bibnamefont{and}
  \bibinfo{author}{\bibfnamefont{A.}~\bibnamefont{{Nunnenkamp}}},
  \bibinfo{journal}{arXiv e-prints} \bibinfo{eid}{arXiv:1910.07539}
  (\bibinfo{year}{2019}), \eprint{1910.07539}.

\bibitem[{\citenamefont{Gambetta
  et~al.}(2019{\natexlab{b}})\citenamefont{Gambetta, Carollo, Lazarides,
  Lesanovsky, and Garrahan}}]{Gambetta2019}
\bibinfo{author}{\bibfnamefont{F.~M.} \bibnamefont{Gambetta}},
  \bibinfo{author}{\bibfnamefont{F.}~\bibnamefont{Carollo}},
  \bibinfo{author}{\bibfnamefont{A.}~\bibnamefont{Lazarides}},
  \bibinfo{author}{\bibfnamefont{I.}~\bibnamefont{Lesanovsky}},
  \bibnamefont{and} \bibinfo{author}{\bibfnamefont{J.~P.}
  \bibnamefont{Garrahan}}, \bibinfo{journal}{Phys. Rev. E}
  \textbf{\bibinfo{volume}{100}}, \bibinfo{pages}{060105}
  (\bibinfo{year}{2019}{\natexlab{b}}),
  \urlprefix\url{https://link.aps.org/doi/10.1103/PhysRevE.100.060105}.

\bibitem[{\citenamefont{{Fan} et~al.}(2019)\citenamefont{{Fan}, {Rossini},
  {Zhang}, {Wu}, {Artoni}, and {La Rocca}}}]{Fan2019}
\bibinfo{author}{\bibfnamefont{C.}~\bibnamefont{{Fan}}},
  \bibinfo{author}{\bibfnamefont{D.}~\bibnamefont{{Rossini}}},
  \bibinfo{author}{\bibfnamefont{H.-X.} \bibnamefont{{Zhang}}},
  \bibinfo{author}{\bibfnamefont{J.-H.} \bibnamefont{{Wu}}},
  \bibinfo{author}{\bibfnamefont{M.}~\bibnamefont{{Artoni}}}, \bibnamefont{and}
  \bibinfo{author}{\bibfnamefont{G.~C.} \bibnamefont{{La Rocca}}},
  \bibinfo{journal}{arXiv e-prints} \bibinfo{eid}{arXiv:1907.03446}
  (\bibinfo{year}{2019}).

\bibitem[{\citenamefont{Matus and Sacha}(2019)}]{Matus2019}
\bibinfo{author}{\bibfnamefont{P.}~\bibnamefont{Matus}} \bibnamefont{and}
  \bibinfo{author}{\bibfnamefont{K.}~\bibnamefont{Sacha}},
  \bibinfo{journal}{Phys. Rev. A} \textbf{\bibinfo{volume}{99}},
  \bibinfo{pages}{033626} (\bibinfo{year}{2019}),
  \urlprefix\url{https://link.aps.org/doi/10.1103/PhysRevA.99.033626}.

\bibitem[{\citenamefont{Zhu et~al.}(2019)\citenamefont{Zhu, Marino, Yao, Lukin,
  and Demler}}]{Zhu2019}
\bibinfo{author}{\bibfnamefont{B.}~\bibnamefont{Zhu}},
  \bibinfo{author}{\bibfnamefont{J.}~\bibnamefont{Marino}},
  \bibinfo{author}{\bibfnamefont{N.~Y.} \bibnamefont{Yao}},
  \bibinfo{author}{\bibfnamefont{M.~D.} \bibnamefont{Lukin}}, \bibnamefont{and}
  \bibinfo{author}{\bibfnamefont{E.~A.} \bibnamefont{Demler}},
  \bibinfo{journal}{New Journal of Physics} \textbf{\bibinfo{volume}{21}},
  \bibinfo{pages}{073028} (\bibinfo{year}{2019}),
  \urlprefix\url{https://doi.org/10.1088/1367-2630/ab2afe}.

\bibitem[{\citenamefont{Bu\v{c}a et~al.}(2019)\citenamefont{Bu\v{c}a, Tindall,
  and Jaksch}}]{Buca2019a}
\bibinfo{author}{\bibfnamefont{B.}~\bibnamefont{Bu\v{c}a}},
  \bibinfo{author}{\bibfnamefont{J.}~\bibnamefont{Tindall}}, \bibnamefont{and}
  \bibinfo{author}{\bibfnamefont{D.}~\bibnamefont{Jaksch}},
  \bibinfo{journal}{Nature Communications} \textbf{\bibinfo{volume}{10}},
  \bibinfo{pages}{1730} (\bibinfo{year}{2019}),
  \urlprefix\url{https://doi.org/10.1038/s41467-019-09757-y}.

\bibitem[{\citenamefont{{Lazarides} et~al.}(2019)\citenamefont{{Lazarides},
  {Roy}, {Piazza}, and {Moessner}}}]{Lazarides2019}
\bibinfo{author}{\bibfnamefont{A.}~\bibnamefont{{Lazarides}}},
  \bibinfo{author}{\bibfnamefont{S.}~\bibnamefont{{Roy}}},
  \bibinfo{author}{\bibfnamefont{F.}~\bibnamefont{{Piazza}}}, \bibnamefont{and}
  \bibinfo{author}{\bibfnamefont{R.}~\bibnamefont{{Moessner}}},
  \bibinfo{journal}{arXiv e-prints} \bibinfo{eid}{arXiv:1904.04820}
  (\bibinfo{year}{2019}), \eprint{1904.04820}.

\bibitem[{\citenamefont{Cai et~al.}(2019)\citenamefont{Cai, huang, and
  Liu}}]{cai2019imaginary}
\bibinfo{author}{\bibfnamefont{Z.}~\bibnamefont{Cai}},
  \bibinfo{author}{\bibfnamefont{Y.}~\bibnamefont{huang}}, \bibnamefont{and}
  \bibinfo{author}{\bibfnamefont{W.~V.} \bibnamefont{Liu}},
  \emph{\bibinfo{title}{Imaginary time crystal of thermal quantum matter}}
  (\bibinfo{year}{2019}), \eprint{1902.09747}.

\bibitem[{\citenamefont{Giergiel et~al.}(2020)\citenamefont{Giergiel, Tran,
  Zaheer, Singh, Sidorov, Sacha, and Hannaford}}]{giergiel2020}
\bibinfo{author}{\bibfnamefont{K.}~\bibnamefont{Giergiel}},
  \bibinfo{author}{\bibfnamefont{T.}~\bibnamefont{Tran}},
  \bibinfo{author}{\bibfnamefont{A.}~\bibnamefont{Zaheer}},
  \bibinfo{author}{\bibfnamefont{A.}~\bibnamefont{Singh}},
  \bibinfo{author}{\bibfnamefont{A.}~\bibnamefont{Sidorov}},
  \bibinfo{author}{\bibfnamefont{K.}~\bibnamefont{Sacha}}, \bibnamefont{and}
  \bibinfo{author}{\bibfnamefont{P.}~\bibnamefont{Hannaford}}
  (\bibinfo{year}{2020}), \eprint{arXiv:2004.00755}.

\bibitem[{\citenamefont{Kuroś et~al.}(2020)\citenamefont{Kuroś, Mukherjee,
  Golletz, Sauvage, Giergiel, Mintert, and Sacha}}]{kuros2020}
\bibinfo{author}{\bibfnamefont{A.}~\bibnamefont{Kuroś}},
  \bibinfo{author}{\bibfnamefont{R.}~\bibnamefont{Mukherjee}},
  \bibinfo{author}{\bibfnamefont{W.}~\bibnamefont{Golletz}},
  \bibinfo{author}{\bibfnamefont{F.}~\bibnamefont{Sauvage}},
  \bibinfo{author}{\bibfnamefont{K.}~\bibnamefont{Giergiel}},
  \bibinfo{author}{\bibfnamefont{F.}~\bibnamefont{Mintert}}, \bibnamefont{and}
  \bibinfo{author}{\bibfnamefont{K.}~\bibnamefont{Sacha}}
  (\bibinfo{year}{2020}), \eprint{arXiv:2004.14982}.

\bibitem[{\citenamefont{Keßler et~al.}(2020)\citenamefont{Keßler, Cosme,
  Georges, Mathey, and Hemmerich}}]{kessler2020}
\bibinfo{author}{\bibfnamefont{H.}~\bibnamefont{Keßler}},
  \bibinfo{author}{\bibfnamefont{J.~G.} \bibnamefont{Cosme}},
  \bibinfo{author}{\bibfnamefont{C.}~\bibnamefont{Georges}},
  \bibinfo{author}{\bibfnamefont{L.}~\bibnamefont{Mathey}}, \bibnamefont{and}
  \bibinfo{author}{\bibfnamefont{A.}~\bibnamefont{Hemmerich}}
  (\bibinfo{year}{2020}), \eprint{arXiv:2004.14633}.

\bibitem[{\citenamefont{Oberreiter et~al.}(2020)\citenamefont{Oberreiter,
  Seifert, and Barato}}]{oberreiter2020}
\bibinfo{author}{\bibfnamefont{L.}~\bibnamefont{Oberreiter}},
  \bibinfo{author}{\bibfnamefont{U.}~\bibnamefont{Seifert}}, \bibnamefont{and}
  \bibinfo{author}{\bibfnamefont{A.~C.} \bibnamefont{Barato}}
  (\bibinfo{year}{2020}), \eprint{arXiv:2002.09078}.

\bibitem[{\citenamefont{Russomanno et~al.}(2020)\citenamefont{Russomanno,
  Notarnicola, Surace, Fazio, Dalmonte, and Heyl}}]{Russomanno2020}
\bibinfo{author}{\bibfnamefont{A.}~\bibnamefont{Russomanno}},
  \bibinfo{author}{\bibfnamefont{S.}~\bibnamefont{Notarnicola}},
  \bibinfo{author}{\bibfnamefont{F.~M.} \bibnamefont{Surace}},
  \bibinfo{author}{\bibfnamefont{R.}~\bibnamefont{Fazio}},
  \bibinfo{author}{\bibfnamefont{M.}~\bibnamefont{Dalmonte}}, \bibnamefont{and}
  \bibinfo{author}{\bibfnamefont{M.}~\bibnamefont{Heyl}},
  \bibinfo{journal}{Phys. Rev. Research} \textbf{\bibinfo{volume}{2}},
  \bibinfo{pages}{012003} (\bibinfo{year}{2020}),
  \urlprefix\url{https://link.aps.org/doi/10.1103/PhysRevResearch.2.012003}.

\bibitem[{\citenamefont{Guo et~al.}(2013)\citenamefont{Guo, Marthaler, and
  Sch\"on}}]{Guo2013}
\bibinfo{author}{\bibfnamefont{L.}~\bibnamefont{Guo}},
  \bibinfo{author}{\bibfnamefont{M.}~\bibnamefont{Marthaler}},
  \bibnamefont{and} \bibinfo{author}{\bibfnamefont{G.}~\bibnamefont{Sch\"on}},
  \bibinfo{journal}{Phys. Rev. Lett.} \textbf{\bibinfo{volume}{111}},
  \bibinfo{pages}{205303} (\bibinfo{year}{2013}),
  \urlprefix\url{https://link.aps.org/doi/10.1103/PhysRevLett.111.205303}.

\bibitem[{\citenamefont{Sacha}(2015{\natexlab{b}})}]{Sacha15a}
\bibinfo{author}{\bibfnamefont{K.}~\bibnamefont{Sacha}}, \bibinfo{journal}{Sci.
  Rep.} \textbf{\bibinfo{volume}{5}}, \bibinfo{pages}{10787}
  (\bibinfo{year}{2015}{\natexlab{b}}),
  \urlprefix\url{https://www.nature.com/articles/srep10787}.

\bibitem[{\citenamefont{Sacha and Delande}(2016)}]{sacha16}
\bibinfo{author}{\bibfnamefont{K.}~\bibnamefont{Sacha}} \bibnamefont{and}
  \bibinfo{author}{\bibfnamefont{D.}~\bibnamefont{Delande}},
  \bibinfo{journal}{Phys. Rev. A} \textbf{\bibinfo{volume}{94}},
  \bibinfo{pages}{023633} (\bibinfo{year}{2016}),
  \urlprefix\url{http://link.aps.org/doi/10.1103/PhysRevA.94.023633}.

\bibitem[{\citenamefont{Guo and Marthaler}(2016)}]{Guo2016}
\bibinfo{author}{\bibfnamefont{L.}~\bibnamefont{Guo}} \bibnamefont{and}
  \bibinfo{author}{\bibfnamefont{M.}~\bibnamefont{Marthaler}},
  \bibinfo{journal}{New Journal of Physics} \textbf{\bibinfo{volume}{18}},
  \bibinfo{pages}{023006} (\bibinfo{year}{2016}),
  \urlprefix\url{http://stacks.iop.org/1367-2630/18/i=2/a=023006}.

\bibitem[{\citenamefont{Guo et~al.}(2016)\citenamefont{Guo, Liu, and
  Marthaler}}]{Guo2016a}
\bibinfo{author}{\bibfnamefont{L.}~\bibnamefont{Guo}},
  \bibinfo{author}{\bibfnamefont{M.}~\bibnamefont{Liu}}, \bibnamefont{and}
  \bibinfo{author}{\bibfnamefont{M.}~\bibnamefont{Marthaler}},
  \bibinfo{journal}{Phys. Rev. A} \textbf{\bibinfo{volume}{93}},
  \bibinfo{pages}{053616} (\bibinfo{year}{2016}),
  \urlprefix\url{https://link.aps.org/doi/10.1103/PhysRevA.93.053616}.

\bibitem[{\citenamefont{Pengfei et~al.}(2018)\citenamefont{Pengfei, Michael,
  and Guo}}]{Liang2017}
\bibinfo{author}{\bibfnamefont{L.}~\bibnamefont{Pengfei}},
  \bibinfo{author}{\bibfnamefont{M.}~\bibnamefont{Michael}}, \bibnamefont{and}
  \bibinfo{author}{\bibfnamefont{L.}~\bibnamefont{Guo}}, \bibinfo{journal}{New
  Journal of Physics} \textbf{\bibinfo{volume}{20}}, \bibinfo{pages}{023043}
  (\bibinfo{year}{2018}), ISSN \bibinfo{issn}{1367-2630},
  \urlprefix\url{http://stacks.iop.org/1367-2630/20/i=2/a=023043}.

\bibitem[{\citenamefont{Giergiel and Sacha}(2017)}]{Giergiel2017}
\bibinfo{author}{\bibfnamefont{K.}~\bibnamefont{Giergiel}} \bibnamefont{and}
  \bibinfo{author}{\bibfnamefont{K.}~\bibnamefont{Sacha}},
  \bibinfo{journal}{Phys. Rev. A} \textbf{\bibinfo{volume}{95}},
  \bibinfo{pages}{063402} (\bibinfo{year}{2017}),
  \urlprefix\url{https://link.aps.org/doi/10.1103/PhysRevA.95.063402}.

\bibitem[{\citenamefont{Mierzejewski et~al.}(2017)\citenamefont{Mierzejewski,
  Giergiel, and Sacha}}]{Mierzejewski2017}
\bibinfo{author}{\bibfnamefont{M.}~\bibnamefont{Mierzejewski}},
  \bibinfo{author}{\bibfnamefont{K.}~\bibnamefont{Giergiel}}, \bibnamefont{and}
  \bibinfo{author}{\bibfnamefont{K.}~\bibnamefont{Sacha}},
  \bibinfo{journal}{Phys. Rev. B} \textbf{\bibinfo{volume}{96}},
  \bibinfo{pages}{140201(R)} (\bibinfo{year}{2017}),
  \urlprefix\url{https://link.aps.org/doi/10.1103/PhysRevB.96.140201}.

\bibitem[{\citenamefont{Delande et~al.}(2017)\citenamefont{Delande,
  Morales-Molina, and Sacha}}]{delande17}
\bibinfo{author}{\bibfnamefont{D.}~\bibnamefont{Delande}},
  \bibinfo{author}{\bibfnamefont{L.}~\bibnamefont{Morales-Molina}},
  \bibnamefont{and} \bibinfo{author}{\bibfnamefont{K.}~\bibnamefont{Sacha}},
  \bibinfo{journal}{Phys. Rev. Lett.} \textbf{\bibinfo{volume}{119}},
  \bibinfo{pages}{230404} (\bibinfo{year}{2017}),
  \urlprefix\url{https://link.aps.org/doi/10.1103/PhysRevLett.119.230404}.

\bibitem[{\citenamefont{Kosior and Sacha}(2018)}]{Kosior2018}
\bibinfo{author}{\bibfnamefont{A.}~\bibnamefont{Kosior}} \bibnamefont{and}
  \bibinfo{author}{\bibfnamefont{K.}~\bibnamefont{Sacha}},
  \bibinfo{journal}{Phys. Rev. A} \textbf{\bibinfo{volume}{97}},
  \bibinfo{pages}{053621} (\bibinfo{year}{2018}),
  \urlprefix\url{https://link.aps.org/doi/10.1103/PhysRevA.97.053621}.

\bibitem[{\citenamefont{Kosior et~al.}(2018)\citenamefont{Kosior, Syrwid, and
  Sacha}}]{Kosior2018a}
\bibinfo{author}{\bibfnamefont{A.}~\bibnamefont{Kosior}},
  \bibinfo{author}{\bibfnamefont{A.}~\bibnamefont{Syrwid}}, \bibnamefont{and}
  \bibinfo{author}{\bibfnamefont{K.}~\bibnamefont{Sacha}},
  \bibinfo{journal}{Phys. Rev. A} \textbf{\bibinfo{volume}{98}},
  \bibinfo{pages}{023612} (\bibinfo{year}{2018}),
  \urlprefix\url{https://link.aps.org/doi/10.1103/PhysRevA.98.023612}.

\bibitem[{\citenamefont{Giergiel
  et~al.}(2018{\natexlab{a}})\citenamefont{Giergiel, Miroszewski, and
  Sacha}}]{Giergiel2018}
\bibinfo{author}{\bibfnamefont{K.}~\bibnamefont{Giergiel}},
  \bibinfo{author}{\bibfnamefont{A.}~\bibnamefont{Miroszewski}},
  \bibnamefont{and} \bibinfo{author}{\bibfnamefont{K.}~\bibnamefont{Sacha}},
  \bibinfo{journal}{Phys. Rev. Lett.} \textbf{\bibinfo{volume}{120}},
  \bibinfo{pages}{140401} (\bibinfo{year}{2018}{\natexlab{a}}),
  \urlprefix\url{https://link.aps.org/doi/10.1103/PhysRevLett.120.140401}.

\bibitem[{\citenamefont{Giergiel
  et~al.}(2018{\natexlab{b}})\citenamefont{Giergiel, Kosior, Hannaford, and
  Sacha}}]{Giergiel2018a}
\bibinfo{author}{\bibfnamefont{K.}~\bibnamefont{Giergiel}},
  \bibinfo{author}{\bibfnamefont{A.}~\bibnamefont{Kosior}},
  \bibinfo{author}{\bibfnamefont{P.}~\bibnamefont{Hannaford}},
  \bibnamefont{and} \bibinfo{author}{\bibfnamefont{K.}~\bibnamefont{Sacha}},
  \bibinfo{journal}{Phys. Rev. A} \textbf{\bibinfo{volume}{98}},
  \bibinfo{pages}{013613} (\bibinfo{year}{2018}{\natexlab{b}}),
  \urlprefix\url{https://link.aps.org/doi/10.1103/PhysRevA.98.013613}.

\bibitem[{\citenamefont{Mizuta et~al.}(2018)\citenamefont{Mizuta, Takasan,
  Nakagawa, and Kawakami}}]{Mizuta2018}
\bibinfo{author}{\bibfnamefont{K.}~\bibnamefont{Mizuta}},
  \bibinfo{author}{\bibfnamefont{K.}~\bibnamefont{Takasan}},
  \bibinfo{author}{\bibfnamefont{M.}~\bibnamefont{Nakagawa}}, \bibnamefont{and}
  \bibinfo{author}{\bibfnamefont{N.}~\bibnamefont{Kawakami}},
  \bibinfo{journal}{Phys. Rev. Lett.} \textbf{\bibinfo{volume}{121}},
  \bibinfo{pages}{093001} (\bibinfo{year}{2018}),
  \urlprefix\url{https://link.aps.org/doi/10.1103/PhysRevLett.121.093001}.

\bibitem[{\citenamefont{Bomantara and Gong}(2018)}]{Bomantara2018}
\bibinfo{author}{\bibfnamefont{R.~W.} \bibnamefont{Bomantara}}
  \bibnamefont{and} \bibinfo{author}{\bibfnamefont{J.}~\bibnamefont{Gong}},
  \bibinfo{journal}{Phys. Rev. Lett.} \textbf{\bibinfo{volume}{120}},
  \bibinfo{pages}{230405} (\bibinfo{year}{2018}),
  \urlprefix\url{https://link.aps.org/doi/10.1103/PhysRevLett.120.230405}.

\bibitem[{\citenamefont{{Khemani} et~al.}(2019)\citenamefont{{Khemani},
  {Moessner}, and {Sondhi}}}]{khemani2019brief}
\bibinfo{author}{\bibfnamefont{V.}~\bibnamefont{{Khemani}}},
  \bibinfo{author}{\bibfnamefont{R.}~\bibnamefont{{Moessner}}},
  \bibnamefont{and} \bibinfo{author}{\bibfnamefont{S.~L.}
  \bibnamefont{{Sondhi}}}, \bibinfo{journal}{arXiv e-prints}
  \bibinfo{eid}{arXiv:1910.10745} (\bibinfo{year}{2019}).

\bibitem[{\citenamefont{{Guo} and {Liang}}(2020)}]{Guo2020}
\bibinfo{author}{\bibfnamefont{L.}~\bibnamefont{{Guo}}} \bibnamefont{and}
  \bibinfo{author}{\bibfnamefont{P.}~\bibnamefont{{Liang}}},
  \bibinfo{journal}{arXiv e-prints} \bibinfo{eid}{arXiv:2005.03138}
  (\bibinfo{year}{2020}).

\bibitem[{\citenamefont{\"Ohberg and Wright}(2019)}]{Ohberg2019}
\bibinfo{author}{\bibfnamefont{P.}~\bibnamefont{\"Ohberg}} \bibnamefont{and}
  \bibinfo{author}{\bibfnamefont{E.~M.} \bibnamefont{Wright}},
  \bibinfo{journal}{Phys. Rev. Lett.} \textbf{\bibinfo{volume}{123}},
  \bibinfo{pages}{250402} (\bibinfo{year}{2019}),
  \urlprefix\url{https://link.aps.org/doi/10.1103/PhysRevLett.123.250402}.

\bibitem[{\citenamefont{Syrwid et~al.}(2020)\citenamefont{Syrwid, Kosior, and
  Sacha}}]{SyrwidKosiorSacha2020}
\bibinfo{author}{\bibfnamefont{A.}~\bibnamefont{Syrwid}},
  \bibinfo{author}{\bibfnamefont{A.}~\bibnamefont{Kosior}}, \bibnamefont{and}
  \bibinfo{author}{\bibfnamefont{K.}~\bibnamefont{Sacha}},
  \bibinfo{journal}{Phys. Rev. Lett.} \textbf{\bibinfo{volume}{124}},
  \bibinfo{pages}{178901} (\bibinfo{year}{2020}),
  \urlprefix\url{https://link.aps.org/doi/10.1103/PhysRevLett.124.178901}.

\bibitem[{\citenamefont{\"{O}hberg and Wright}(2020)}]{ReplyOhbergWright2020}
\bibinfo{author}{\bibfnamefont{P.}~\bibnamefont{\"{O}hberg}} \bibnamefont{and}
  \bibinfo{author}{\bibfnamefont{E.~M.} \bibnamefont{Wright}},
  \bibinfo{journal}{Phys. Rev. Lett.} \textbf{\bibinfo{volume}{124}},
  \bibinfo{pages}{178902} (\bibinfo{year}{2020}),
  \urlprefix\url{https://link.aps.org/doi/10.1103/PhysRevLett.124.178902}.

\bibitem[{\citenamefont{Castin}(2001)}]{Castin_LesHouches}
\bibinfo{author}{\bibfnamefont{Y.}~\bibnamefont{Castin}}, in
  \emph{\bibinfo{booktitle}{Coherent atomic matter waves}}, edited by
  \bibinfo{editor}{\bibfnamefont{R.}~\bibnamefont{Kaiser}},
  \bibinfo{editor}{\bibfnamefont{C.}~\bibnamefont{Westbrook}},
  \bibnamefont{and} \bibinfo{editor}{\bibfnamefont{F.}~\bibnamefont{David}}
  (\bibinfo{publisher}{Springer Berlin Heidelberg}, \bibinfo{address}{Berlin,
  Heidelberg}, \bibinfo{year}{2001}), pp. \bibinfo{pages}{1--136}, ISBN
  \bibinfo{isbn}{978-3-540-45338-3}.

\bibitem[{\citenamefont{Pethick and Smith}(2002)}]{Pethick2002}
\bibinfo{author}{\bibfnamefont{C.}~\bibnamefont{Pethick}} \bibnamefont{and}
  \bibinfo{author}{\bibfnamefont{H.}~\bibnamefont{Smith}},
  \emph{\bibinfo{title}{{Bose-Eistein condensation in dilute gases}}}
  (\bibinfo{publisher}{{Cambridge University Press}},
  \bibinfo{address}{{Cambridge, England}}, \bibinfo{year}{2002}).

\bibitem[{\citenamefont{Carr et~al.}(2000)\citenamefont{Carr, Clark, and
  Reinhardt}}]{Carr2000}
\bibinfo{author}{\bibfnamefont{L.~D.} \bibnamefont{Carr}},
  \bibinfo{author}{\bibfnamefont{C.~W.} \bibnamefont{Clark}}, \bibnamefont{and}
  \bibinfo{author}{\bibfnamefont{W.~P.} \bibnamefont{Reinhardt}},
  \bibinfo{journal}{Phys. Rev. A} \textbf{\bibinfo{volume}{62}},
  \bibinfo{pages}{063611} (\bibinfo{year}{2000}),
  \urlprefix\url{http://link.aps.org/doi/10.1103/PhysRevA.62.063611}.

\bibitem[{\citenamefont{Kanamoto
  et~al.}(2003{\natexlab{a}})\citenamefont{Kanamoto, Saito, and
  Ueda}}]{Kanamoto2003}
\bibinfo{author}{\bibfnamefont{R.}~\bibnamefont{Kanamoto}},
  \bibinfo{author}{\bibfnamefont{H.}~\bibnamefont{Saito}}, \bibnamefont{and}
  \bibinfo{author}{\bibfnamefont{M.}~\bibnamefont{Ueda}},
  \bibinfo{journal}{Phys. Rev. A} \textbf{\bibinfo{volume}{67}},
  \bibinfo{pages}{013608} (\bibinfo{year}{2003}{\natexlab{a}}),
  \urlprefix\url{https://link.aps.org/doi/10.1103/PhysRevA.67.013608}.

\bibitem[{\citenamefont{Kanamoto
  et~al.}(2003{\natexlab{b}})\citenamefont{Kanamoto, Saito, and
  Ueda}}]{Kanamoto2003B}
\bibinfo{author}{\bibfnamefont{R.}~\bibnamefont{Kanamoto}},
  \bibinfo{author}{\bibfnamefont{H.}~\bibnamefont{Saito}}, \bibnamefont{and}
  \bibinfo{author}{\bibfnamefont{M.}~\bibnamefont{Ueda}},
  \bibinfo{journal}{Phys. Rev. A} \textbf{\bibinfo{volume}{68}},
  \bibinfo{pages}{043619} (\bibinfo{year}{2003}{\natexlab{b}}),
  \urlprefix\url{https://link.aps.org/doi/10.1103/PhysRevA.68.043619}.

\bibitem[{\citenamefont{Kanamoto et~al.}(2009)\citenamefont{Kanamoto, Carr, and
  Ueda}}]{Kanamoto2009}
\bibinfo{author}{\bibfnamefont{R.}~\bibnamefont{Kanamoto}},
  \bibinfo{author}{\bibfnamefont{L.~D.} \bibnamefont{Carr}}, \bibnamefont{and}
  \bibinfo{author}{\bibfnamefont{M.}~\bibnamefont{Ueda}},
  \bibinfo{journal}{Phys. Rev. A} \textbf{\bibinfo{volume}{79}},
  \bibinfo{pages}{063616} (\bibinfo{year}{2009}),
  \urlprefix\url{https://link.aps.org/doi/10.1103/PhysRevA.79.063616}.

\bibitem[{\citenamefont{Edmonds et~al.}(2013)\citenamefont{Edmonds, Valiente,
  Juzeliunas, Santos, and \"Ohberg}}]{Edmonds2013}
\bibinfo{author}{\bibfnamefont{M.~J.} \bibnamefont{Edmonds}},
  \bibinfo{author}{\bibfnamefont{M.}~\bibnamefont{Valiente}},
  \bibinfo{author}{\bibfnamefont{G.}~\bibnamefont{Juzeliunas}},
  \bibinfo{author}{\bibfnamefont{L.}~\bibnamefont{Santos}}, \bibnamefont{and}
  \bibinfo{author}{\bibfnamefont{P.}~\bibnamefont{\"Ohberg}},
  \bibinfo{journal}{Phys. Rev. Lett.} \textbf{\bibinfo{volume}{110}},
  \bibinfo{pages}{085301} (\bibinfo{year}{2013}),
  \urlprefix\url{https://link.aps.org/doi/10.1103/PhysRevLett.110.085301}.

\bibitem[{\citenamefont{Lehtovaara et~al.}(2007)\citenamefont{Lehtovaara,
  Toivanen, and Eloranta}}]{Lehtovaara2007}
\bibinfo{author}{\bibfnamefont{L.}~\bibnamefont{Lehtovaara}},
  \bibinfo{author}{\bibfnamefont{J.}~\bibnamefont{Toivanen}}, \bibnamefont{and}
  \bibinfo{author}{\bibfnamefont{J.}~\bibnamefont{Eloranta}},
  \bibinfo{journal}{J. of Comp. Phys.} \textbf{\bibinfo{volume}{221}},
  \bibinfo{pages}{148} (\bibinfo{year}{2007}),
  \urlprefix\url{https://doi.org/10.1016/j.jcp.2006.06.006}.

\bibitem[{q_f()}]{q_footnote}
\bibinfo{note}{Note that $g_c$ does not change when we change $q$ within the
  same parity class. Such a feature manifests at the level of the GPE,
  Eq.~(\ref{GPE_W}), and energy functional, Eq.~(\ref{elab_CS}). Indeed, if
  $\Psi$ is a solution of Eq.~(\ref{GPE_W}) with $q=q_0$, then $\Psi_s=\Psi
  e^{i s \theta}$ is a solution of the same problem but with $q=q_0+2s$ where
  $s\in\mathbb{Z}$, i.e. the change $q\rightarrow q+2s$ only modifies the phase
  winding numbers of the GPE solutions so that $J\rightarrow J+s$.
  Additionally, while the energies, Eq.~(\ref{elab_CS}), corresponding to
  $\Psi$ and $\Psi_s$ differ only by a constant associated with the change
  $W=q^2/4\rightarrow(q+2s)^2/4$, the probability current, Eq.~(\ref{probCur}),
  remains unchanged.}

\end{thebibliography}


\end{document}